# Genetic transfer in *Staphylococcus*: a case study of 13 genomes

Cheong Xin Chan[1], Robert G. Beiko[2] and Mark A. Ragan[1]

[1]ARC Centre of Excellence in Bioinformatics and Institute for Molecular Bioscience, The University of Queensland, Brisbane, QLD 4072, Australia

[2]Department of Computer Science, Dalhousie University, 6050 University Avenue, Halifax, Nova Scotia, Canada B3H 1W5

## Abstract

The widespread presence of antibiotic resistance and virulence among *Staphylococcus* isolates has been attributed to lateral genetic transfer (LGT) between different strains or species. However, there has been very little study of the extent of LGT in *Staphylococcus* species using a phylogenetic approach, particularly of the units of such genetic transfer. Here we report the first systematic study of the units of genetic transfer in 13 *Staphylococcus* genomes, using a rigorous phylogenetic approach. We found clear evidence of LGT in 26.1% of the 1354 homologous gene families examined, and possibly more in another 17.9% of the total families. Within-gene and whole-gene transfer contribute almost equally to the discordance of these gene families against a reference phylogeny. Comparing genetic transfer in single-copy and in multi-copy gene families, we found little functional bias in cases of within-gene (fragmentary) genetic transfer but substantial functional bias in cases of whole-gene (non-fragmentary) genetic transfer, and we observed a higher frequency of LGT in multi-copy gene families. Our results demonstrate that LGT and gene duplication play an important part among the factors that contribute to functional innovation in staphylococcal genomes.

## Introduction

*Staphylococcus* species are a group of non-motile but invasive Gram-positive bacteria that have been associated to various pus-forming diseases in humans and other animals. *Staphylococcus aureus* is the most prominent pathogenic species in the genus, in which a variety of strains are found to colonise the nasal passages and skin in human, causing diverse illnesses that range from minor skin lesions or infections to life-threatening diseases *e.g.* meningitis, septicaemia and Toxic Shock Syndrome [1-3]. The other species of *Staphylococcus*, although lacking genes that encode virulence factors and toxins, have been found to cause a number of opportunistic infections in immune-compromised patients [4].

One of the major problems in the prognosis of staphylococcal infections is the development of resistance in the bacteria to multiple antibiotics over the years. Penicillin was initially used for treatment of infections caused by *S. aureus* in 1940s, but a number of strains soon developed resistance to such antibiotics [5, 6]. Subsequent introduction of other antibiotics similar to penicillin, *e.g.* streptomycin, tetracycline and chloramphenicol, was also counteracted by the emergence of staphylococcal strains that are resistant; the beta-lactamase produced by these organisms can break down the beta-lactam ring, a core structure in penicillin and its related antibiotics [7]. Consequently, semi-synthetic antibiotics that are resistant to beta-lactamase (*e.g.* methicillin) and antibiotics with no beta-lactam ring (*e.g.* vancomycin) were developed, but isolates of *S. aureus* that are resistant to both types of antibiotics have been reported in recent years [8-12], suggesting that the list of antibiotics that can be used to treat staphylococcal infections is near exhaustion [13, 14].

The widespread presence of antibiotic resistance in *Staphyloccoccus* species has been attributed to the susceptibility of the organisms to genetic transfer, in which the organisms acquire exogenous genetic material that codes for antibiotic resistance and maintain these sequences in the genome by selection [15]. In general, bacteria can share genes between different species or strains through lateral genetic transfer (LGT) *via* the mechanisms of transduction, transformation and/or conjugation. The exogenous genetic material can be integrated into the recipient genome through genetic recombination. In *Staphylococcus*, phage-mediated conjugation is one of the more common mechanisms of genetic transfer, in which the presence of bacteriophages can increase the adhesiveness of the bacterial cell surface and therefore assist in the conjugative transfer of genetic materials between two organisms [16].

The genomic makeup of each strain or species of *Staphylococcus* can occur in the form of a single core genome, or as a core genome that co-exists with one or more accessory genomes *e.g.* plasmids, mobile genetic elements and pathogenicity islands. In a number of multiple loci sequence typing (MLST) studies examining variations of highly conserved housekeeping genes, the genes in the core genomes were found to conform largely to vertical inheritance [17-19]. Genes in the accessory genomes, however, were found to have arisen largely by genetic transfer and recombination events [14, 19]). A number of such genes are associated with virulence factors, *e.g.* 'superantigen' genes within pathogenicity islands that are implicated in toxic shock and food poisoning [14]. These studies demonstrate that genetic transfer, particularly LGT, is a major factor that contributes to the evolution of virulence and antibiotic resistance in *Staphylococcus* species.

While a number of studies have examined the frequency of LGT in prokaryotes using rigorous phylogenetic approaches [20, 21], studies of LGT in *Staphylococcus* have been limited. The extent of LGT in *Staphylococcus* species was first explored based on simple similarity matches among gene sequences [22]. A rigorous phylogenetic approach was adopted in a recent study to compare the phylogenies of putatively orthologous gene families from eight *Staphylococcus* genomes to a reference species phylogeny [23], in which pathogenicity-related and extra-chromosomal genes were found to be likely to have arisen *via* LGT. These studies have been based on the implicit assumption that the units of genetic transfer are whole genes; within-gene (fragmentary) genetic transfer was not considered. It has been shown, however, that the genetic material integrated in the LGT process can constitute an entire gene [24], a partial (fragmentary) gene [25, 26], or multiple (entire or fragmentary) adjacent genes [27, 28]. Studies that focus entirely on whole-gene transfer can underestimate the extent of LGT.

We previously examined the units of genetic transfer in 144 prokaryote genomes using a comprehensive rigorous phylogenetic approach [29]. In that study, we focused on non-duplicated genes from each genome to avoid the complications of paralogy in the inference of LGT. Here we report on the units of genetic transfer across 13 fully sequenced genomes of *Staphylococus*, with the inclusion of genes that have been duplicated in each genome. Within a gene family, the presence of two or more homologous gene copies from the same genome can be interpreted as paralogs (a result of within-genome duplication) or xenologs (a result from acquisition of a gene copy from an external source *via* LGT). Since we do not have prior knowledge on the origins of such multiple gene copies, herein we follow Lerat *et al.* in referring to them as *synologs* [21]. We characterise the frequencies of within- and whole-gene transfer in gene families that contain no synolog and those that contain one or more synologs, and discuss correlations with annotated gene functions. This represents the first systematic study on the units of genetic transfer in prokaryotes using a rigorous phylogenetic approach, in which duplicated gene copies have been considered.

# Results

We analysed the units of genetic transfer in 1354 protein-coding gene families from 13 fully sequenced genomes of *Staphylococcus* isolates. The 13 genomes represent four different species: *S. aureus* (9 isolates), *S. epidermidis* (2 isolates), *S. haemophyticus* (1 isolate) and *S. saprophyticus* (1 isolate). The representation of each genome in the 1354 gene families is shown in Figure 1. The only virulent species, *S. aureus*, is well-represented, with all but two *S. aureus* genomes represented in at least 1200 gene families.

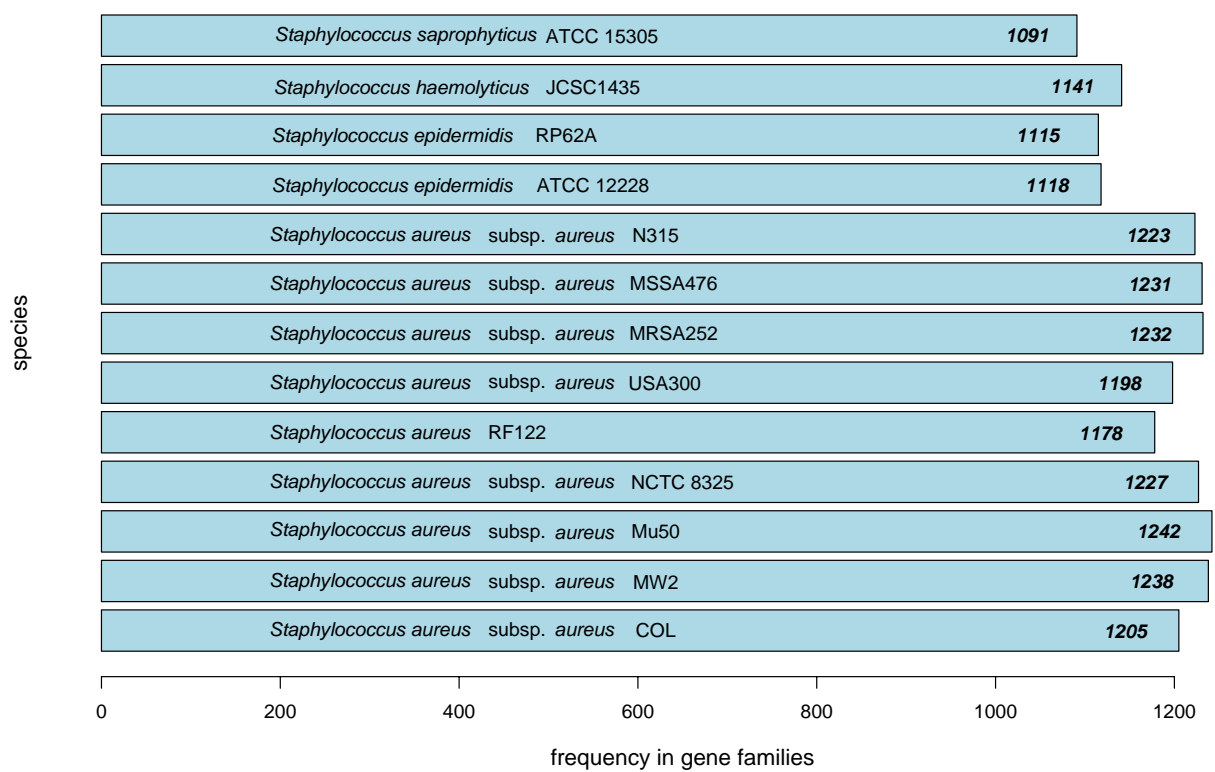

Figure 1. Representation of the 13 *Staphylococcus* strains within 1354 gene families.

The genomes of *S. aureus* subsp. *aureus* N315 and Mu50 were the first of any *Staphylococcus* to be sequenced and released [22]; both strains were isolated from male patients with post-surgical wound infections in Japan. The strains MW2 [30] and MSSA476 [31] are both isolates of community-acquired *S. aureus* infections; the strain COL [4] is the oldest isolate of methicillin-resistant *S. aureus* (MRSA) dating back to 1976; the strain

USA300 [32] was isolated from unassociated outbreaks of *S. aureus* infections in USA, Canada and Europe within the last decade; and the strain NCTC8325 [33] has been used as the generic representative strain of *S. aureus* in most genetic studies. The *S. aureus* strain MRSA252 [31] is a hospital-acquired MRSA, while RF122 [34] is a common strain associated with mastitis diseases in cattle. Two-thirds of the strains of *S. aureus* included in this study are methicillin-resistant, *i.e.* all except MSSA476, NCTC8325 and RF122.

The other three species, *S. epidermidis*, *S. haemolyticus* and *S. saprophyticus*, are non-virulent but have been associated with a number of opportunistic infections in immune-compromised patients. *S. epidermidis* strain RP62A [4] is a pathogenic, slime-producing strain that can cause Toxic Shock Syndrome and scarlet fever, whereas the strain ATCC1228 [35] of the same species is a non-biofilm forming and non-pathogenic strain that has been used for detecting residual antibiotics in food products. *S. haemolyticus* (strain JCSC1435) [36] and *S. saprophyticus* (strain ATCC 15305) [37] are generally opportunistic pathogens; *S. haemolyticus* infrequently causes soft tissue infections, and *S. saprophyticus* is predominantly implicated in genitourinary tract infections.

The protein-coding sequences were initially clustered into 2924 protein families, 77 (2.6%) of which are of size $N > 52$. Many of these large families consist of proteins related to transport functions, including the largest family, with 1777 proteins related to the ATP-binding cassette (ABC) transporter, one of the largest protein families known in prokaryotes [38]. The second-largest family is a 769-member cluster of proteins related to the phototransferase system (PTS), which is involved in sugar phosphorylation and regulation of metabolic processes [39]. Families with $N < 4$ do not contribute to meaningful phylogenetic inference, and phylogenetic inference for families with a large $N$ is not only computationally intensive, but will be difficult to interpret with only 13 distinct isolates. We therefore restricted the dataset to gene families with $4 \leq N \leq 52$ after removal of identical sequences (see Materials and Methods). This yielded 1354 gene families with $4 \leq N \leq 41$, as shown in Figure 2 (no family fell in the range $N = 42\text{-}52$). A total of 229 (15.7%) gene families are of size $N = 8$, and 197 (14.5%) are of $N = 9$. The largest family in the dataset consists of 41 sequences that encode for putative nucleotidase proteins, while the second largest family has 39 putative proteins of guanosine- or inositol-monophosphate dehydrogenase.

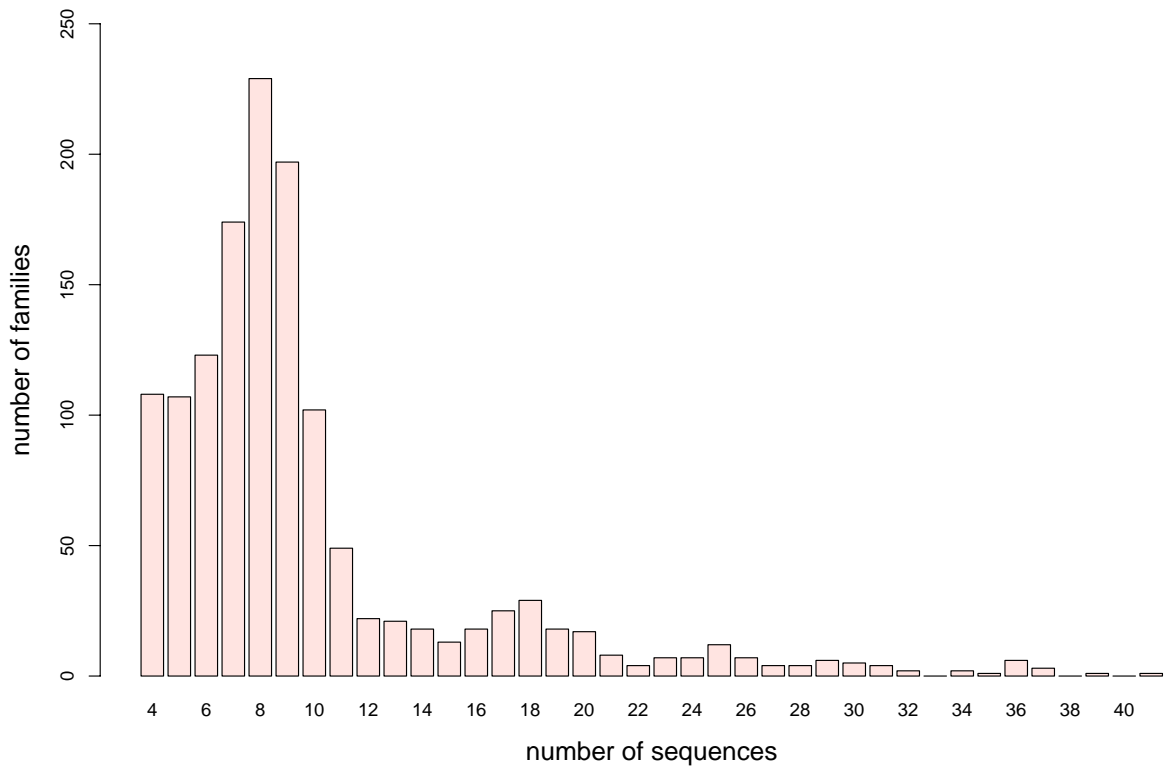

Figure 2. Size distribution of gene families examined in this study.

*Inference based on evidence of genetic transfer*

We examined the evidence of genetic transfer in the 1354 gene families, focusing on the units of such transfer in single- and multi-copy gene families. *Single-copy* gene families contain no within-genome duplicated genes, *i.e.* no genome is represented more than once in such a family. In the absence of evidence to the contrary, we assume that these genes share a common evolutionary origin prior to the divergence of these strains, *i.e.* to be orthologs. On the other hand, *multi-copy* gene families contain one or more within-genome duplicates, and at least one genome is represented more than once in such a family. The additional gene copy, whichever it is (it is not necessary to distinguish the "original" from the duplicate), must either have arisen *via* a within-genome duplication event (*i.e.* be a paralog) or have been imported *via* LGT into the genome (*i.e.* be a xenolog). In the absence of prior or external knowledge of their origin, we call them synologs [21]. The presence or absence of synologs necessarily affects how we interpret evidence of genetic transfer, as discussed below.

Evidence of recombination (phylogenetic discrepancy) and inference of a breakpoint within the boundaries of the gene family is interpreted as *within-gene* (fragmentary) genetic transfer. Table 1 shows the possible interpretations of fragmentary transfer in single-copy (*sc*) and multi-copy (*mc*) gene families.

Table 1. Inference based on evidence of within-gene transfer in single-copy (*sc*) and multi-copy (*mc*) gene families, for cases with negative evidence (*neg*) and positive evidence inferred as LGT (*lgt*). In the cases of multi-copy gene families, the suffix *P* denotes that synologs are paralogs, and *X* denotes that synologs are xenologs. *P/X* denotes that each synolog is either a paralog or a xenolog (but not both), while *PX* denotes that each synolog can have a complex history, *i.e.* be a paralog, a xenolog, or both. See text for details.

|  | **Evidence of within-gene transfer** | | | |
| --- | --- | --- | --- | --- |
|  | **Negative** | **Positive** | | |
| (Single-copy gene families (*sc*) | No LGT<br>*sc-neg* | Within-gene LGT between orthologs<br>*sc-lgt* | | |
| Multi-copy gene families (*mc*) | No LGT<br>*mc-neg-P* | No synologs are recombinant<br>(synolog = paralog)<br>*mc-lgt-P* | Some synologs are recombinant<br>(synolog = paralog or xenolog)<br>*mc-lgt-P/X* | All synologs are recombinant<br>(synolog = paralog and/or xenolog)<br>*mc-lgt-PX* |

For those families in which we find no internal recombination breakpoint, we infer that there has been no within-gene transfer (*sc-neg* and *mc-neg-P*); in the latter case (multi-copy families), all synologs are paralogs (*P*). We have already argued that phylogenetic discordance and discovery of a recombination breakpoint implies within-gene lateral genetic transfer (*lgt*). In single-copy families, such evidence (*sc-lgt*) can be interpreted (again, in the absence of evidence to the contrary) as LGT between orthologs. In multi-copy families, however, the situation is more complex: (a) if LGT is inferred within the family but subtrees that include synologs are not topologically incongruent with the reference tree (*i.e.* recombination has affected only sequences other than the synologs), then all synologs are paralogs (*mc-lgt-P*). (b) Where some but not all of the synologs are recombinant, the non-recombinant synologs are native paralogs but the recombinants are xenologs (or more precisely, paralog-xenolog chimeras), and we denote this situation in the family by *mc-lgt-P/X*. Where per-genome copy number is small, it is reasonable to assume (in the absence of evidence to the contrary) that the LGT event occurred subsequent to the within-genome duplication. (c) If, however, all synologs are detected as recombinant, we know that each is a xenolog (again, more precisely, a paralog-xenolog chimera); but it may be impossible to reconstruct the precise order of within-genome duplication, LGT and perhaps other (*e.g.* gene conversion or lineage-sorting) events that have produced this situation, which we denote *mc-lgt-PX*.

The interpretation is much the same in the case of *whole-gene* transfer, as shown in Table 2 for single-copy (*SC*) and multi-copy (*MC*) gene families, although with a complication (discussed below) that arises from our methodological approach. We use capital letters to distinguish these families from those affected by fragmentary transfer (Table 1). In the case of single-gene families, both alternatives (*SC-neg* and *SC-LGT*) exactly parallel those for within-gene transfer, with whole-gene LGT in single-copy families again interpreted as involving orthologs. In multi-copy families, the situation is again more complex: (a) if LGT is inferred within the family but the synologs are not implicated, then all synologs are paralogs (*MC-LGT-P*). (b) Where some but not all of the synologs are recombinant, the non-recombinant synologs are native paralogs but the recombinants are xenologs (not chimeras, as these genes have been transferred in their entirety), and we denote this situation in the family by *MC-LGT-P/X* with the same qualification as in the case of within-gene transfer. (c) If all synologs are recombinant, we infer that each is a xenolog throughout its coding region, although as before it may be impossible to reconstruct the precise order of within-genome duplication, LGT and perhaps other events that have produced this situation, which we denote *MC-LGT-PX*.

Table 2. Inference based on evidence of whole-gene transfer in single-copy (*SC*) and multi-copy (*MC*) gene families, for cases with negative evidence (*neg*) and positive evidence inferred as LGT (*LGT*). The capital letters in the labels distinguish these cases from the within-gene transfer shown in Table 1. Labels otherwise follow the conventions introduced in Table 1. See text for details.

|  | Evidence of whole-gene transfer | | | |
| --- | --- | --- | --- | --- |
|  | **Negative** | **Positive** | | |
| Single-copy gene families (*SC*) | No LGT<br>*SC-neg* | Whole-gene LGT involving orthologs<br>*SC-LGT* | | |
| Multi-copy gene families (*MC*) | No LGT<br>*MC-neg-P* | No recombining sequences are synologs<br>(synolog = paralog)<br>*MC-LGT-P* | Some recombining sequences are synologs<br>(synolog = paralog or xenolog)<br>*MC-LGT-P/X* | All recombining sequences are synologs<br>(synolog = paralog and/or xenolog)<br>*MC-LGT-PX* |

The complication, mentioned above, concerns the evidentiary basis on which we identify none, some, or all synologs as recombinant *via* whole-gene transfer. In our approach, families for which evidence of genetic transfer is detected in the first phase (see Methods and Materials) are examined more rigorously to identify recombination breakpoint(s). The software (DualBrothers [40]) we found optimal for this latter task works by identifying the point(s) at which phylogenetic trees inferred for the sequence regions to the immediate left and right disagree. These trees can be recovered and compared with the reference supertree, making it straightforward to identify recombinant region(s) within-gene. Where no breakpoint can be identified, it is necessary to infer trees in a separate step for comparison against the reference topology. As a general rule we carried this out as just described, *i.e.* by using the synologous gene family as input into the Bayesian inference software MRBAYES [41]. Two drawbacks of this approach are the resource (CPU and memory) demands associated with inference from large datasets, and the complexity of extracting and interpreting topology data for all minimal subtrees that include synologs, particularly when taking into account the relative support for relevant bipartitions.

> We realised that there is an opportunity to address both of these drawbacks for the subset of families (about one-third: see later) that contain only a few (here: one or two) synologs. For each such gene family, we de-replicated synologs in all combinations, yielding a set of alignments in which each genome is represented only once; see

Figure 3 for an example. Then from each alignment we inferred a Bayesian phylogenetic tree. As each family contains either one or two synologs, for each we generate either two or four trees, each of which we compare separately against the reference supertree and can easily automate the extraction of the relevant congruence relationship. This approach, adapted from [21], could in principle be extended to greater numbers of synologs per family, although at exponential cost.

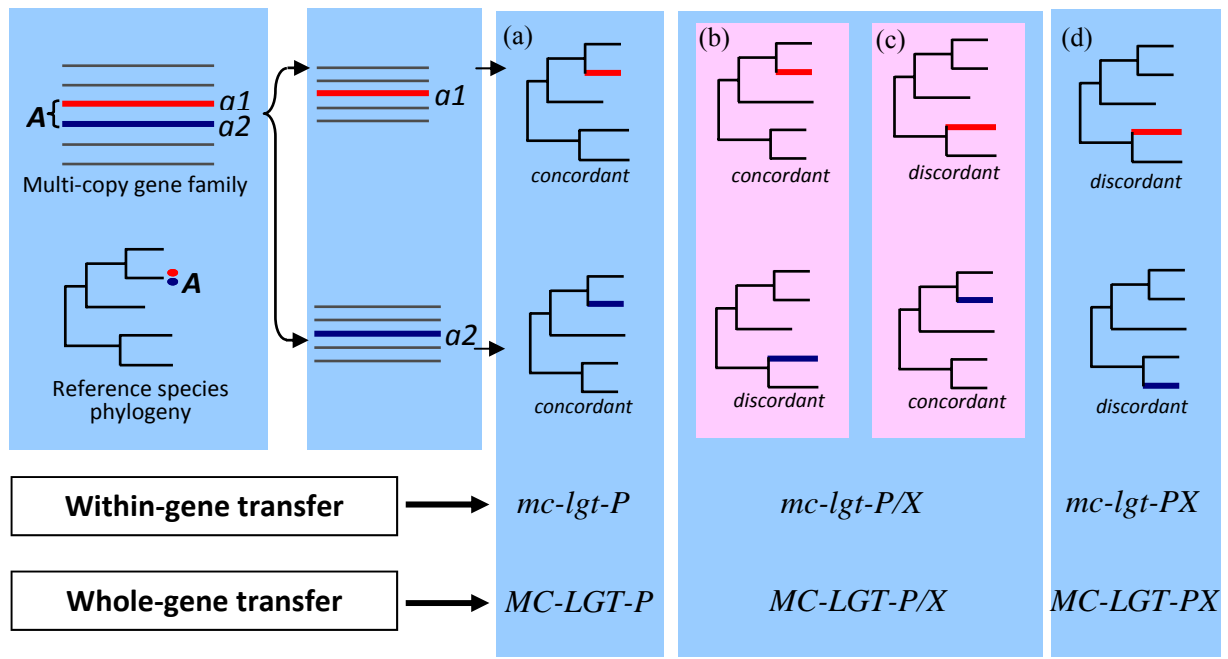

Figure 3. Testing the source of synologs in multi-copy gene families. The illustrated example is a family with a single synolog, in which genome *A* has two gene copies, *a1* and *a2*. The synologs were de-replicated in all combinations, yielding two sets of alignments in which each genome is represented only once. A Bayesian phylogenetic tree was constructed for each of these alignments, and the tree was compared against the reference species phylogeny. There are four possible outcomes in such a comparison with respect to the reference phylogeny: (a) both trees of the de-replicated alignments are concordant, suggesting both *a1* and *a2* are paralogs; in (b) and (c), either one of the trees is concordant while the other is discordant, suggesting each of *a1* and *a2* is either a paralog or a xenolog (but not both); and (d) both trees are discordant, suggesting both *a1* and *a2* have a complex history, *i.e.* be a paralog, a xenolog or both. The labels for each inference of synolog are shown for multi-copy gene families (*mc*), in cases of within-gene transfer (small letters) and whole-gene transfer (all capital letters).

*Within-gene (fragmentary) genetic transfer*

We applied a two-phase strategy [42] to detect recombination in each of the 1354 gene families. Three statistical tests [43-45] were used in the first phase to detect the occurrences of recombination based on discrepancies in phylogenetic signals within each gene family, and recombination was inferred when at least two of the three tests show a *p*-value ≤ 0.10. A rigorous Bayesian phylogenetic approach, as implemented in DualBrothers [40], was applied in the second phase to locate the recombination breakpoints more precisely in gene families that showed evidence of recombination in the first phase. The inferred recombination within these gene families represents the evidence of fragmentary genetic transfer, as one or more breakpoints that separate the sequence set into two topologically distinct regions are present within the gene families.

Of the 1354 gene families, we found 401 (29.6%) that show evidence of within-family recombination after the first phase screening. Of these families, 252 (18.6% of 1354) show clear evidence of recombination based on Bayesian phylogenetic analysis in the second phase, with Bayesian posterior probability (BPP) support ≥ 0.500 for the dominant topology on at least one side of the inferred breakpoint in each family. For 68 families (5.0%) we could identify a recombination breakpoint, but no sequence region has BPP ≥ 0.500; we labelled these as inconclusive. For a further 81 families (6.0%), recombination was detected in the first phase but no recombination breakpoint could be identified in the second phase. No evidence of recombination was detected in 953 (70.4%) gene families in the first-phase screening.

Of the 252 gene families that show clear evidence of within-gene (fragmentary) genetic transfer, 98 families are single-copy gene families (*sc-lgt*) while the other 154 are multi-copy gene families (*mc-lgt*). Phylogenetic discrepancy at different regions across a gene-family alignment can sometimes be due to regional (*e.g.* domain-specific) differences in rates of nucleotide substitution in one or more sequences. We looked specifically for such differential rates ($\mu \geq 0.30$, from DualBrothers) but did not observe any instances that span the inferred breakpoints. This suggests that the breakpoints inferred herein indeed arise from genetic recombination.

To examine possible functional bias pertaining to fragmentary genetic transfer within the 252 families, we used annotations from the TIGR Comprehensive Microbial Resource (http://cmr.tigr.org/) to assign a functional category (TIGR role category) to each family.

Figure 4 shows the proportions of proteins in each functional category for (a) single-copy (*sc-lgt*) and (b) multi-copy (*mc-lgt*) gene families for which we inferred fragmentary genetic transfer, compared to their frequencies in the full (1354-family) staphylococcal dataset.

Families affected by fragmentary genetic transfer are significantly either over- or under-represented in more than half of the TIGR role categories, and this is the case for both single- and multi-copy gene families. Families affected by fragmentary transfer are significantly over-represented, compared with expectation, in the energy metabolism, DNA metabolism, protein fate, amino acid biosynthesis, nucleotide synthesis, and transcription categories for both *sc-lgt* and *mc-lgt*. On the other hand, both types of families are significantly under-represented in the hypothetical protein, transport and binding, regulatory-function categories. Families involved in protein synthesis are over-represented in *sc-lgt* but under-represented in *mc-lgt*, while protein families of unknown function are biased in the opposite direction.

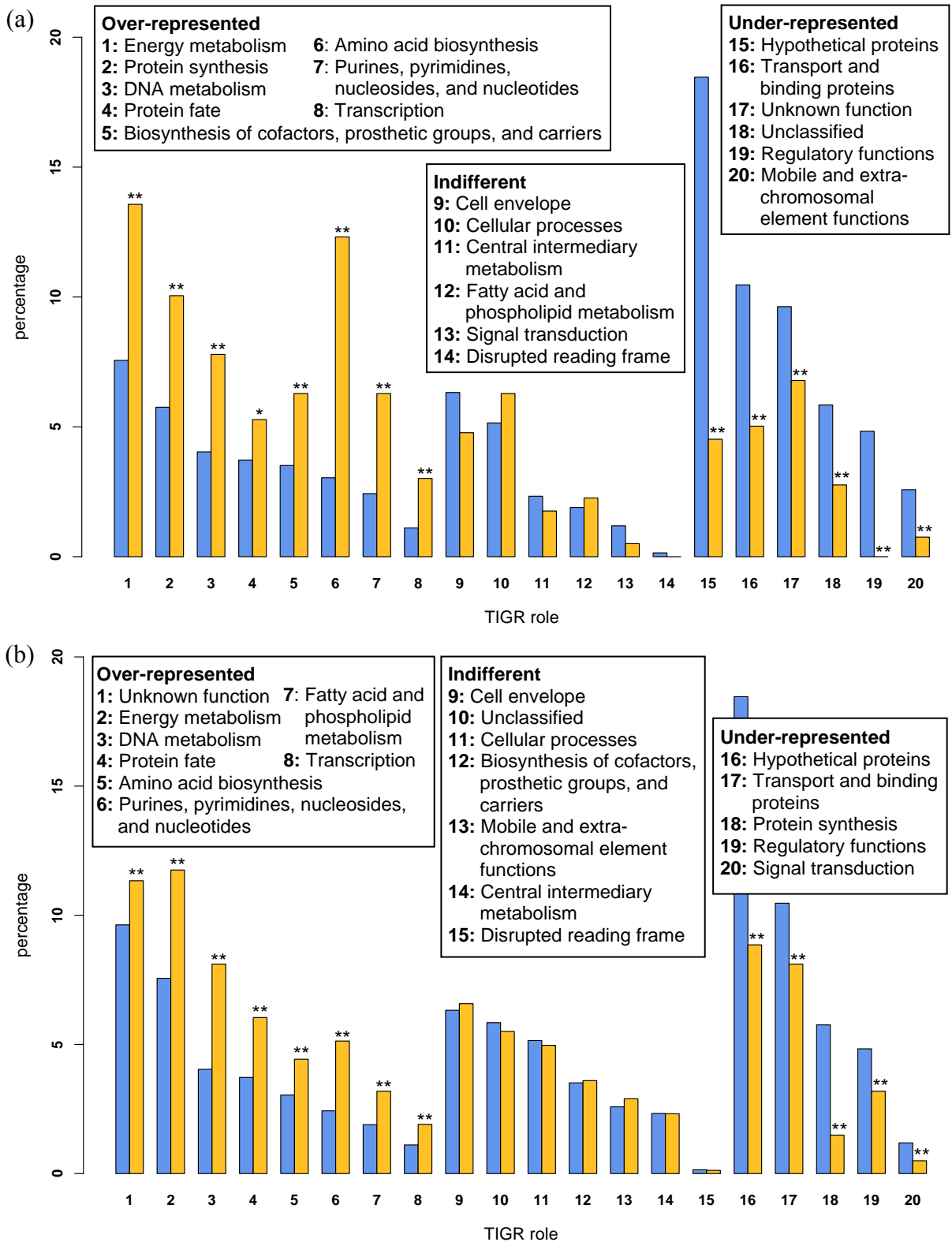

Figure 4. Representation of functional categories assigned to protein sequences corresponding to gene families in *Staphylococcus* species that show evidence of within-gene (fragmentary) genetic transfer (yellow bars) for (a) single-copy gene families, *sc-lgt* and (b) multi-copy gene families, *mc-lgt*. The blue bars show the representation of these same functional categories in the full dataset (1354 families, 13297 proteins). Categories are numbered differently for panels (a) and (b) as shown in the boxes. Significance of over- or under-representation is represented by single ($p \leq 0.05$) and double asterisks ($p \leq 0.01$).

*Whole-gene (non-fragmentary) genetic transfer*

Of the 1354 staphylococcal gene families under consideration, we have so far inferred within-gene transfer for 320 (252 as clear instances of LGT, and a further 68 for which evidence was deemed inconclusive). We now turn to the 1034 remaining families: 953 recombination-negative from the first phase, plus 81 recombination-positive from the first phase but breakpoint-negative from the second. For each we inferred a Bayesian phylogenetic tree, and compared it with the reference supertree generated from 2645 putatively orthologous families in these 13 genomes using the method of matrix representation with parsimony (MRP) [46]; see Materials and Methods for further details. The MRP supertree is shown in Figure 5. Four species of *Staphylococcus* species are represented among the 13 genomes, each monophyletic according to our analysis. Whole-gene recombination was inferred in a gene family if the tree topology is discordant with that of the reference tree.

The 1034 gene families can be divided into three classes based on the number of synologs present in each family. The number of gene families with trees concordant or discordant with the reference supertree for each category is shown in Table 3.

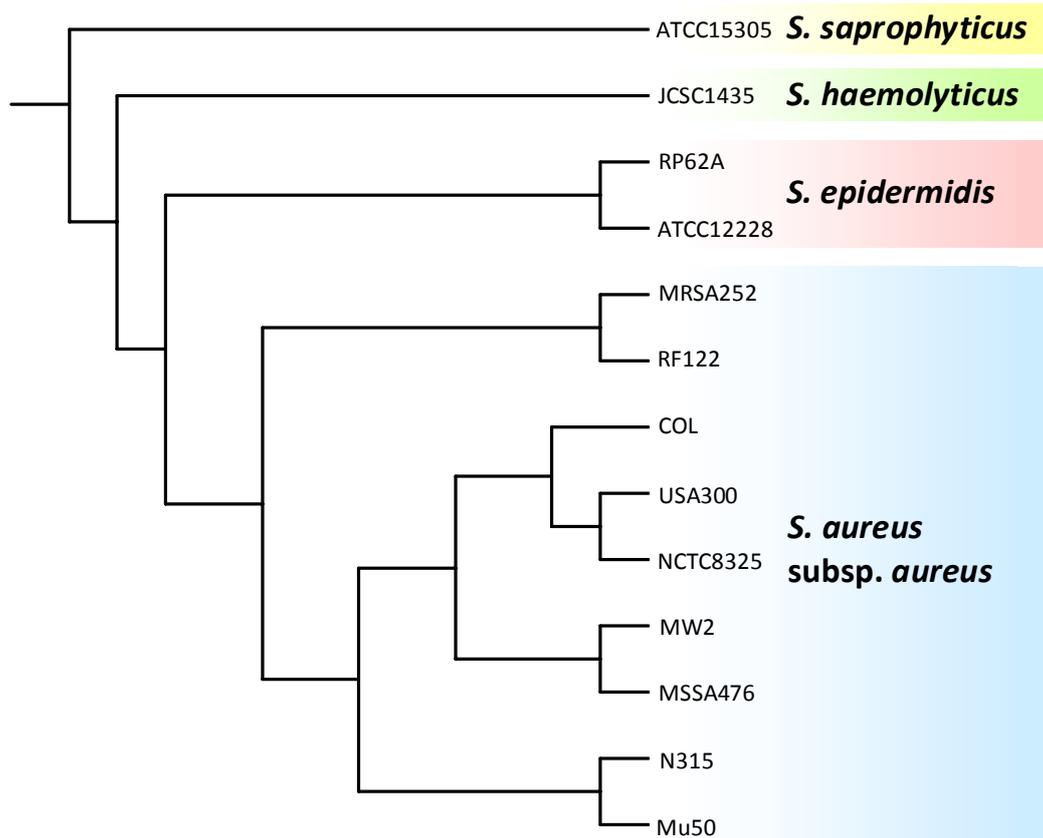

Figure 5. Reference supertree for the 13 staphylococcal genomes, rooted based on previous phylogenetic study on *Staphylococcus* species using small subunit rRNA genes [47] that gives *S. saprophyticus* as the outgroup.

Table 3. Counts of inferred whole-gene transfer based on topological comparison between the phylogenetic tree inferred for each gene family and the reference supertree. The frequency of concordance and discordance in these families, based on four maximum-likelihood tests, are shown; proportions are relative to the total dataset of 1354 gene families. Labels of different categories follow the description in Table 2.

| Category | Tree comparison based on maximum-likelihood tests | |
| --- | --- | --- |
| | Concordant | Discordant |
| Single-copy gene families, no synologs (*SC*) | 684<br>*SC-neg* | 90<br>*SC-LGT* |
| Families with one or two synologs each (within *MC*) | 54<br>*MC-neg-P* | 38<br>*MC-LGT* |
| Families with more than two synologs each (within *MC*) | 14<br>*MC-neg-P* | 154<br>*MC-LGT* |
| **Total number of families (%)** | **752 (55.5%)** | **282 (20.8%)** |

Among the 774 of these families in *SC*, 90 (12%) show topological discordance and represent LGT involving orthologs, whereas no evidence of LGT was found among the other 684 (88%). Among the 260 of these families in *MC*, however, the outcome was very different: 192 (74%) are discordant *vis-à-vis* the reference topology, and only 68 (26%) concordant. Of these 260 families, 92 contain one or two synologs each, and 168 contain more than two. Of the 92 families in the former group, 38 (41%) revealed evidence of whole-gene LGT whereas, remarkably, among the 168 in the latter group, fully 154 (92%) did so. Using our synolog de-replication approach (

Figure 3), we could further classify the 38 *MC-LGT* families with one or two synologs into five for which all synologs are paralogs but not xenologs (*MC-LGT-P*), 21 for which synologs are either paralogs or xenologs but not both (*MC-LGT-P/X*), and 12 with more-complex histories (*MC-LGT-PX*).

Figure 6 shows the proportions of proteins in each TIGR role functional category within (a) single-copy (*SC-LGT*) and (b) multi-copy (*MC-LGT*) gene families for which we inferred whole-gene transfer, compared to their frequencies in the full staphylococcal dataset. The proteins in these families are significantly either over- or under-represented in more than half of the TIGR role categories, although with more-numerous and greater differences between single- and multi-copy gene families than was observed for fragmentary transfer. Single-copy families affected by whole-gene transfer are significantly over-represented, compared with expectation, in the protein synthesis, central intermediary metabolism and transcription categories are significantly over-represented, suggesting that these *Staphylococcus* genomes appear to have been more receptive to introgression of whole genes that encode for these protein functions when no indigenous copy was already present in the genome, or if an indigenous copy was present it was replaced or subsequently lost.

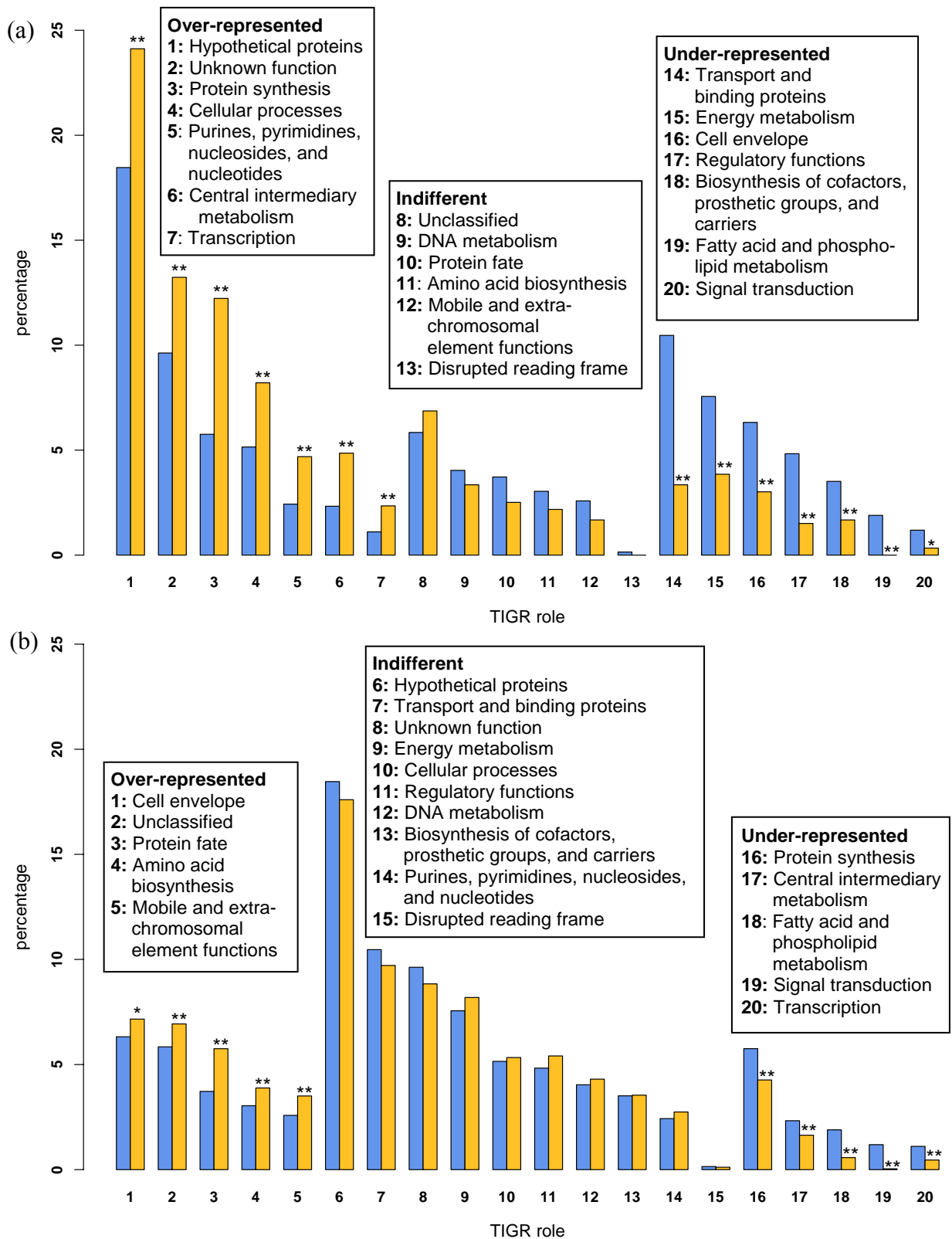

Figure 6. Representation of functional categories assigned to protein sequences corresponding to gene families in *Staphylococcus* species that show evidence of whole-gene transfer (yellow bars) for (a) single-copy gene families, *SC-LGT* and (b) multi-copy gene families, *MC-LGT*. The blue bars show the representation of these same functional categories in the full dataset (1354 families, 13297 proteins). Categories are numbered differently for panels (a) and (b) shown in the boxes. Significance of over- or under-representation is represented by single ($p \leq 0.05$) and double asterisks ($p \leq 0.01$).

Correspondingly, families in these same functional categories are significantly under-represented in *MC-LGT*, suggesting that these genomes have been less receptive to the integration of genes encoding these functions when multiple copies already exist. Families engaged in cellular processes are over-represented in *SC-LGT*, particularly those involved in toxin production and resistance ($p = 1.7 \times 10^{-4}$), while families involved in complex regulatory functions including energy metabolism are under-represented. Families involved in cell envelope functions, and those related to mobile and extra-chromosomal genetic elements, are over-represented in *MC-LGT*. Families implicated in fatty acid metabolism and signal transduction are under-represented in both *SC-LGT* and *MC-LGT*.

*Correlation of transferred genomic regions with protein structural domains*

Within each gene family for which we found evidence of within-gene (fragmentary) genetic transfer, we further asked whether the transferred coding regions correlate with protein structural domains. We obtained domain information for each protein by sequence similarity matching to entries in the Pfam [48] and SCOP [49] databases. For each inferred recombination breakpoint, we computed a breakpoint-to-boundary distance that represents the distance (in amino acid residues) between an inferred breakpoint and the nearest annotated domain boundary. If the transferred gene regions are correlated with domain structure, inferred breakpoints will be located close to domain boundaries (*i.e.* the breakpoint-to-boundary distance will be smaller than would be expected under a model that incorporates no correlation). The distribution of observed distances was compared against a distribution of expected distances generated *via* permutation (see Materials and Methods). Figure 7 shows the breakpoint-to-boundary distances obtained for (a) single-copy gene families, *i.e. sc-lgt* and (b) multi-copy gene families, *i.e. mc-lgt*. For each, we show the (i) observed distances,

(ii) expected distances, and (iii) a quantile-quantile plot between the observed and expected distances.

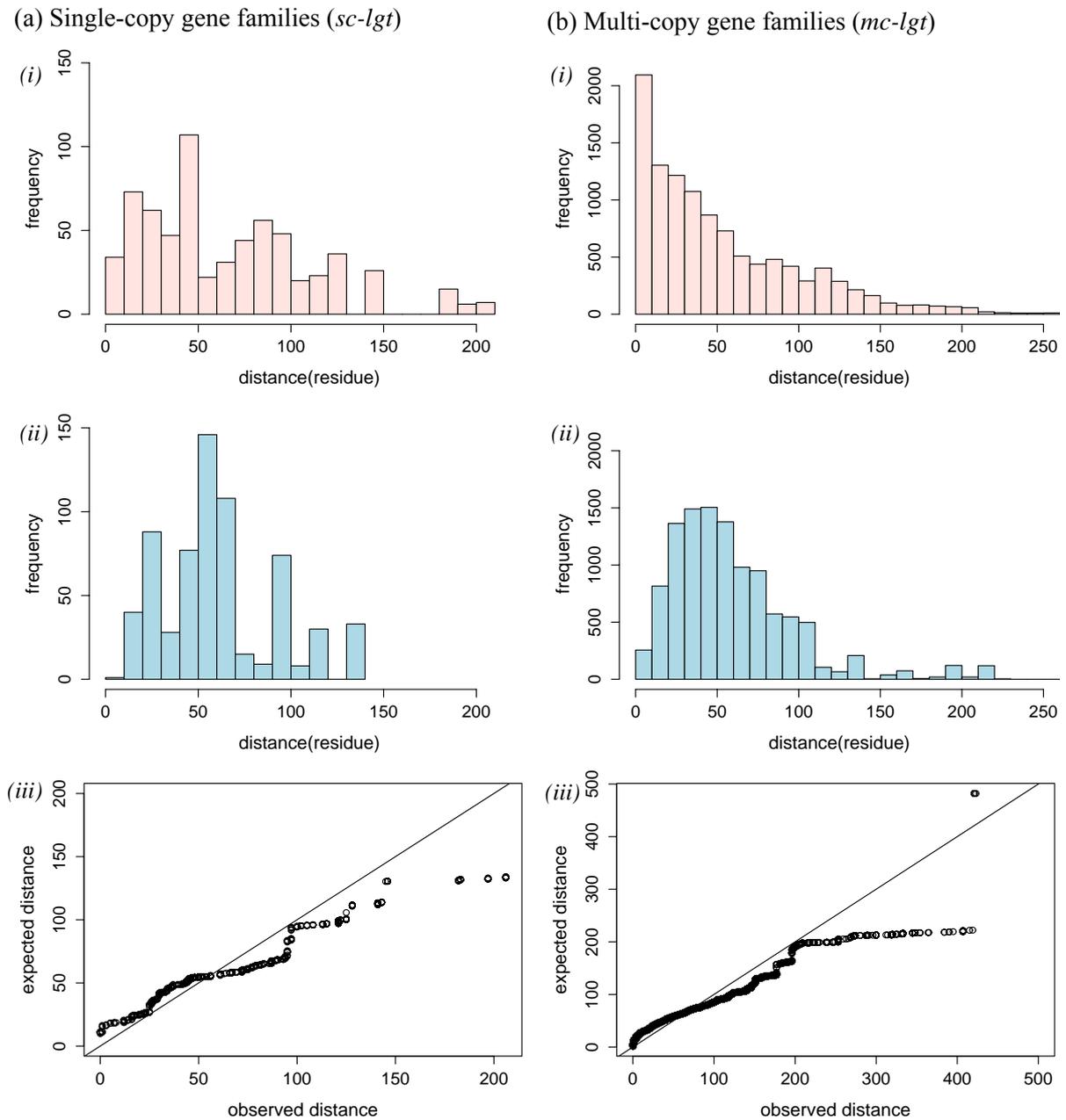

Figure 7. Distances between inferred breakpoint and the nearest protein domain boundary in (a) single-copy gene families (*sc-lgt*) and (b) multi-copy gene families (*mc-lgt*) in these *Staphylococcus* genomes. (*i*) Observed distances, (*ii*) expected distances averaged from 100000 permuted breakpoints, and (*iii*) the quantile-quantile plot between the observed and expected distances, are shown for each of (a) and (b), respectively.

For both *sc-lgt* and *mc-lgt*, we find minimal but significant differences between observed and expected distances (Kolmogorov-Smirnov test, *D* values < 0.2, $p < 10^{-10}$), and the observed distances are shorter than expected up to about 50 amino acid residues (Figures 6a(iii) and 6b(iii)). For *mc-lgt*, although not all breakpoint-to-boundary distances are significantly closer than expected, 41.5% of the inferred breakpoints are ≤ 30 amino acid residues from the nearest domain boundary.

**Discussion**

Most phylogenetic approaches to quantifying LGT in prokaryotes [20, 21, 50], including a recent study on eight genomes of *Staphylococcus* [23], have been limited by the intrinsic assumption that the unit of genetic transfer is a whole gene. Discordance of topology between a gene-family tree and the reference tree has been interpreted as *prima facie* evidence that a gene has been transferred from one lineage into another. We recently reported a rigorous phylogenetic approach in which we examined the extent of LGT in prokaryotes without restricting the unit of transfer to be a whole gene [29], and found, within a set of single-copy gene families in 144 prokaryote genomes [20], that within-gene genetic transfer is about twice as frequent as the transfer of entire genes (or larger units). By adopting a similar approach to 1354 gene families from 13 *Staphylococcus* genomes in the current study, we found clear evidence of LGT in 354 families (252 in *sc-lgt,* 90 in *SC-LGT,* 12 in *MC-LGT-PX*; 26.1% of 1354), and probably more in another 243 families (68 inconclusive, 175 in *MC-LGT-P/X*; 17.9% of 1354), in which the origins of synologs cannot be determined. Within-gene (18.6%) and whole-gene (20.8%) transfer contribute almost equally to the discordance of the gene families against the reference species phylogeny. We observed a higher frequency of genetic transfer involving multi-copy than single-copy gene families, some of which can be explained by LGT alone, but most of which may reflect more-complex evolutionary histories involving (for example) multiple LGT and gene duplication events, gene conversion and/or lineage sorting. On the other hand, our approach cannot detect recombination between genomic lineages that are terminal and adjacent on a given gene-family tree, and to the extent such transfer has actually taken place, our estimates are necessarily low by the corresponding amount.

LGT and gene duplication have been proposed to be the major contributing factors of functional innovations in the genomes of prokaryotes [51]. Through LGT, advantageous genes that are not otherwise present within a lineage can be integrated into the genome and contribute to the organism's fitness, while gene duplication can relax selective constraints, allowing gene products to explore specialised functions. Following duplication, a genetic region can lose its original function (non-functionalisation), gain a novel function (neofunctionalisation), or take on a specialised part of the original function (subfunctionalisation) [52]. Hooper and Berg [53] proposed that duplication is more common among laterally transferred genes than among indigenous ones. Based on the results presented here we cannot reject this assertion, as we found a large number of multi-copy gene families that show evidence of whole-gene LGT.

Families for which we infer LGT are represented at frequencies significantly different from expectation in more than half of these TIGR role (functional) categories. Their patterns of over- and under-representation differ more between single-copy and multi-copy families for whole-gene than for fragmentary LGT. Families involved in protein synthesis and affected by LGT, whether within- or whole-gene, are very significantly over-represented among single-copy families but very significantly under-represented among multi-copy families, suggesting that these *Staphylococcus* genomes are more susceptible to introgression *via* LGT of genes encoding protein-synthetic enzymes when no similar copy is already present in the recipient genome than when multiple copies already exist. Biases of this magnitude between the two types of families are also seen for the TIGR role categories of cell envelope, central intermediary metabolism, and transcription in the case of whole-gene transfer, and for proteins of unknown function in the case of fragmentary transfer. Our results demonstrate that some genes that encode informational proteins are susceptible to both within-gene and whole-gene LGT in *Staphylococcus*; the introgression of these genetic fragments into the bacterial genome could be an important repair process to preserve gene function.

For whole-gene transfer we likewise found significant bias in favour of single-gene families, in the frequency of protein families engaged in toxin production and resistance. This result suggests that the broad distribution of antibiotic resistance across the different isolates of *Staphylococcus* [15] might be explained, at least in part, by the tendency of these organisms to incorporate these genes in their entirety *via* LGT when no similar sequences already existed in the recipient genome.

We previously reported, in 144 prokaryote genomes, that the locations of recombination breakpoints are correlated with protein domain boundaries [54]. Here, for within-gene genetic transfer in *Staphylococcus* genomes, we found weak (although significant) correlation between inferred breakpoints and annotated protein domain boundaries. The relative weakness of the correlation, compared to what we observed earlier, can probably be attributed to the similarity of the sequences in the *Staphylococcus* dataset, which comprises 13 *Staphylococcus* genomes including nine isolates of *S. aureus*. The overall pairwise nucleotide sequence similarity across all gene families is high, ranging from 0.47 to 0.99 (mean 0.74 ± 0.12 s.d.). In the previous study [54] we showed that correlation between breakpoint locations and domain boundaries is lower when sequence similarity is high (nucleotide identity > 0.5). The reason for this may be twofold: highly similar sequences are unlikely to be disruptive of protein domain boundaries, even when the recombination breakpoint is distant; and breakpoints become increasingly difficult to locate precisely at high similarity between introgressed and native sequences, as characteristic differences necessarily become less-frequent.

The events of LGT and genetic duplication (*i.e.* xenologous and paralogous genes) are two contributing factors that shape the functional evolution of certain protein families in prokaryotes [55]. Our results demonstrate that both of these events also play an important part among the factors that contribute to functional innovation in staphylococcal genomes. Nevertheless, the extent of genetic transfer in these genomes might be underestimated in the current study, as complicated evolutionary history with multiple events of genetic transfer and/or genetic duplication, *e.g.* recombination between two synologs or duplication of synologs, may not have been possible to detect with the available methodology.

## Materials and Methods

### Data

The thirteen fully sequenced genomes of *Staphylococcus* as of 7 September 2006 were downloaded from GenBank (http://www.ncbi.nlm.nih.gov/), as shown in Table 4.

All protein-coding sequences (34066) were clustered into 2924 protein families based on similarity matching *via* a Markov clustering algorithm (inflation parameter 1.1) [56]. These families consist of homologous protein sequences; some include one or more copies of genes, *i.e.* synologs. Multiple sequence alignment was performed on each protein family using T-COFFEE [57] with four combinations of gap opening (25, 50) and gap extension (0, 10) penalties, and using MUSCLE [58] with default settings. The alignments were validated using a pattern-centric objective function [59], in which the alignment that was assigned the highest score by the objective function was selected as the optimal alignment for each protein family.

Table 4. Thirteen genomes of *Staphylococcus* used in this study, with their respective RefSeq accession numbers and the total number of coding sequences.

| Genome (*Staphylococcus* sp.) | RefSeq accession | Number of coding sequences |
| --- | --- | --- |
| *S. aureus* RF122 | NC_007622 | 2515 |
| *S. aureus* subsp. *aureus* COL | NC_002951 | 2618 |
| *S. aureus* subsp. *aureus* MRSA252 | NC_002952 | 2656 |
| *S. aureus* subsp. *aureus* MSSA476 | NC_002953 | 2598 |
| *S. aureus* subsp. *aureus* Mu50 | NC_002758 | 2731 |
| *S. aureus* subsp. *aureus* MW2 | NC_003923 | 2632 |
| *S. aureus* subsp. *aureus* N315 | NC_002745 | 2619 |
| *S. aureus* subsp. *aureus* NCTC8325 | NC_007795 | 2892 |
| *S. aureus* subsp. *aureus* USA300 | NC_007793 | 2604 |
| *S. epidermidis* ATCC12228 | NC_004461 | 2485 |
| *S. epidermidis* RP62A | NC_002976 | 2526 |
| *S. haemolyticus* JCSC1435 | NC_007168 | 2676 |
| *S. saprophyticus* subsp. *saprophyticus* ATCC 15305 | NC_007350 | 2514 |

The protein alignments were converted into nucleotide sequence alignments, in which the nucleotide triplets were arranged to parallel exactly the protein alignment in each case. For these nucleotide sequence alignments, we require family size $N \geq 4$ because 4 is the minimum size that can yield distinct topologies if every sequence in the family is unique. We identified sets of identical nucleotide sequences and removed (at random) all but one copy of each, after which families with $N < 4$ (1493, 51.1%) were excluded from the analysis. Almost all of the identical copies (99.9%) removed from the dataset represented organisms from the same species or strain. In addition, families with $N > 52$ (77, 2.6%) were excluded, since inference of phylogeny on such large families, most of which (56, 1.9%) have $N > 70$, would create problems both with inference and with interpretation. In this way, our final dataset used was reduced to 1354 families ($4 \leq N \leq 52$) encompassing 13297 genes (39.0% of all genes available in these genomes).

*Reference species tree*

As reference phylogenetic tree for the 13 species we computed a supertree using matrix representation with parsimony [46] from 2645 putatively orthologous protein families generated from all 34066 protein-coding sequences *via* a hybrid clustering approach [60]. The distances among the species were calculated using CLANN version 2.0.1 [61] before matrix and tree reconstruction using PAUP* version 4.0 [62].

*Detecting within-gene (fragmentary) genetic transfer*

A two-phase strategy [42] was used for detecting recombination in the gene families. Three statistics, maximal chi-squared [44], neighbour similarity score [45] and pairwise homoplasy index implemented in PhiPack [43], were first used to detect evidence of recombination events within the sequence sets based on discrepancies in phylogenetic signals. Each test yields a *p*-value indicating the statistical significance of such phylogenetic discrepancy, *i.e.* the presence of a recombination event. Datasets with at least two of the three *p*-values $\leq 0.10$ were considered as positives. Once evidence of recombination was identified in a gene family, we used DualBrothers [40] to define the breakpoints more precisely. The program was run with MCMC chain length = 2500000, burnin = 500000, window_length = 5 and Peter Green's constant C = 0.25. The MCMC search space was determined separately for each gene family, in which the search space represents a list of all possible phylogenetic trees that could be inferred from shorter segments within the sequence set. MRBAYES (MCMC chain length = 2500000, burnin = 500000, nucmodel = 4by4, rates=gamma, ngammacat = 4) was used to infer unrooted phylogenetic trees for each partition (window) that was slid across the alignment (window length = 100, sliding size = 50; unit in alignment position). The inferred trees within the threshold at Bayesian Confidence Interval (BCI) of 90% were included in the initial tree list, with the maximum set at 1000. Gene families that show evidence of recombination are inferred to have undergone one or more events of fragmentary genetic transfer.

*Detecting whole-gene (non-fragmentary) genetic transfer*

For each gene family in which no evidence of recombination was found in the first-phase screening, and for those positive in the first-phase screening but in which no recombination breakpoint could be detected, we inferred a Bayesian phylogenetic tree and compared its topology against that of the species reference tree. Whole-gene (non-fragmentary) genetic transfer was inferred if the topologies were significantly discordant. The approach for analysis of genetic transfer in families that contain one or more synologs is adopted from a previous study [21]. Each gene family containing *m* number of genes that are represented by *n* number of genomes was categorised into three separate classes: (a) those with no synolog, $m = n$; (b) those with one or two synologs, $m > n$ and $m \leq n + 2$; and (c) those with more than two synologs, $m > n + 2$. For (a) families with no synologs, topological discordance between a gene tree and the reference supertree implies a whole-gene genetic transfer *via* LGT. For each family in (b), two or four new alignment sets were generated by removing replicate synologs in all combinations, then inferring and comparing trees. Discordance between one or more of these gene trees and the reference supertree implies acquisition of a synolog in the family *via* LGT. For (c) families with more than two synologs, generation and analysis of de-replicated trees in all possible combinations quickly becomes impractical. For these families, we compared the gene tree directly to the reference supertree; discordance suggests whole-gene recombination and might be explained by one or more events as discussed in the text.

Individual gene-family trees were inferred from DNA alignments using MRBAYES [41] with MCMC chain length = 2500000, burnin = 500000, and model = K2P [63]. The phylogeny of each gene family was compared against the reference supertree using four maximum-likelihood tests at 95% confidence level: the Shimodaira-Hasegawa test [64], the one- and two-sided Kishino-Hasegawa tests [65, 66], and expected likelihood weights [67], all as implemented in Tree-Puzzle 5.1 [68]. Discordance between the two trees was inferred when significant exclusion of the reference tree was indicated in more than two of the four tests. Such discordance was taken as *prima facie* evidence of whole-gene (non-fragmentary) genetic transfer.

*Functional analysis of gene families*

Functional information for each protein sequence was retrieved from the Comprehensive Microbial Resource (CMR) at The Institute for Genomic Research website (http://cmr.tigr.org/), based on TIGR role identifiers and categorisation at Level 1. Over- or under-representation of functional categories was based on the probability of observing a defined number of target groups (or categories) in a subsample, given a process of sampling without replacement from the whole dataset (as defined in each case: see text) under a hypergeometric distribution [69]. The probability of observing *x* number of a particular target category is described as:

$$P(k=x) = f(k;N,m,n) = \frac{\binom{m}{k}\binom{N-m}{n-k}}{\binom{N}{n}}$$

in which *N* is the total population size, *m* is the size of the target category within the population, *n* is the total size of the subsample, and *k* is the size of the target category within the subsample.

*Correlation between breakpoint location and protein domain boundary*

Protein domain and boundary information for each protein in the dataset ($N = 13297$) was determined by sequence similarity search against domain entries (type = 'domain') in Pfam version 20 [48] and SCOP version 1.69 [49]. The breakpoint-to-boundary distance, adapted from previous studies [70, 71], is defined as the distance from an inferred recombination breakpoint to the nearest annotated domain boundary, in the unit of amino acid residues. The expected breakpoint-to-boundary distance was obtained by averaging the distances from a series 100000 permutations. Each permutation was carried out by assigning a random recombination breakpoint in the sequence and calculating the distance of the breakpoint to the closest domain boundary that is annotated on the sequence. The distributions of observed and expected distances were then compared using Kolmogorov-Smirnov test [72].

## Acknowledgements

This study was supported by Australian Research Council (ARC) grant CE0348221. We thank Aaron Darling and Vladimir Minin for valuable advice on the use of DualBrothers. CXC was supported by a University of Queensland UQIPRS scholarship.

## References

1. Lowy FD (1998). *Staphylococcus aureus* infections. *N. Engl. J. Med.*, **339**: 520-532.

2. Kluytmans J, van Belkum A and Verbrugh H (1997). Nasal carriage of *Staphylococcus aureus*: epidemiology, underlying mechanisms, and associated risks. *Clin. Microbiol. Rev.*, **10**: 505-520.

3.  Lyon BR and Skurray R (1987). Antimicrobial resistance of *Staphylococcus aureus*: genetic basis. *Microbiol Rev*, **51**: 88-134.

4.  Gill SR, Fouts DE, Archer GL, Mongodin EF, Deboy RT, Ravel J, Paulsen IT, Kolonay JF, Brinkac L, Beanan M, Dodson RJ, Daugherty SC, Madupu R, Angiuoli SV, Durkin AS, Haft DH, Vamathevan J, Khouri H, Utterback T, Lee C, Dimitrov G, Jiang L, Qin H, Weidman J, Tran K, Kang K, Hance IR, Nelson KE and Fraser CM (2005). Insights on evolution of virulence and resistance from the complete genome analysis of an early methicillin-resistant *Staphylococcus aureus* strain and a biofilm-producing methicillin-resistant *Staphylococcus epidermidis* strain. *J. Bacteriol.*, **187**: 2426-2438.

5.  North EA and Christie R (1946). Acquired resistance of staphylococci to the action of penicillin. *Med. J. Aust.*, **1**: 176-179.

6.  Barber M (1947). Staphylococcal infection due to penicillin-resistant strains. *BMJ*, **2**: 863-865.

7.  Piorde JJ and Sherris JC (1974). Staphylococcal resistance to antibiotics: origin, measurement, and epidemiology. *Ann. N. Y. Acad. Sci.*, **236**: 413-434.

8.  Fridkin SK, Hageman JC, Morrison M, Sanza LT, Como-Sabetti K, Jernigan JA, Harriman K, Harrison LH, Lynfield R and Farley MM (2005). Methicillin-resistant *Staphylococcus aureus* disease in three communities. *N. Engl. J. Med.*, **352**: 1436-1444.

9.  Chambers HF (1997). Methicillin-resistance in staphylococci: molecular and biochemical basis and clinical implications. *Clin. Microbiol. Rev.*, **10**: 781-791.

10. Deresinski S (2005). Methicillin-resistant *Staphylococcus aureus*: an evolutionary, epidemiologic, and therapeutic odyssey. *Clin. Infect. Dis.*, **40**: 562-573.

11. Hiramatsu K (2001). Vancomycin-resistant *Staphylococcus aureus*: a new model of antibiotic resistance. *Lancet Infect Dis*, **1**: 147-155.

12. Tenover FC, Weigel LM, Appelbaum PC, McDougal LK, Chaitram J, McAllister S, Clark N, Killgore G, O'Hara CM, Jevitt L, Patel JB and Bozdogan B (2004). Vancomycin-resistant *Staphylococcus aureus* isolate from a patient in Pennsylvania. *Antimicrob. Agents Chemother.*, **48**: 275-280.

13. Foster TJ (2004). The *Staphylococcus aureus* 'superbug'. *J Clin Invest*, **114**: 1693-1696.

14. Lindsay JA and Holden MTG (2004). *Staphylococcus aureus*: superbug, super genome? *Trends Microbiol.*, **12**: 378-385.

15. Grundmann H, Aires-de-Sousa M, Boyce J and Tiemersma E (2006). Emergence and resurgence of meticillin-resistant *Staphylococcus aureus* as a public-health threat. *Lancet*, **368**: 874-885.

16. Lacey RW (1980). Evidence for two mechanisms of plasmid transfer in mixed cultures of *Staphylococcus aureus*. *J. Gen. Microbiol.*, **119**: 423-435.

17. Feil EJ and Spratt BG (2001). Recombination and the population structures of bacterial pathogens. *Annu. Rev. Microbiol.*, **55**: 561-590.

18. Harmsen D, Claus H, Witte W, Rothganger J, Turnwald D and Vogel U (2003). Typing of methicillin-resistant *Staphylococcus aureus* in a university hospital setting by using novel software for *spa* repeat determination and database management. *J. Clin. Microbiol.*, **41**: 5442-5448.

19. Hiramatsu K, Watanabe S, Takeuchi F, Ito T and Baba T (2004). Genetic characterization of methicillin-resistant *Staphylococcus aureus*. *Vaccine*, **22**: S5-S8.

20. Beiko RG, Harlow TJ and Ragan MA (2005). Highways of gene sharing in prokaryotes. *Proc. Natl. Acad. Sci. U. S. A.*, **102**: 14332-14337.

21. Lerat E, Daubin V, Ochman H and Moran NA (2005). Evolutionary origins of genomic repertoires in bacteria. *PLoS Biol.*, **3**: e130.

22. Kuroda M, Ohta T, Uchiyama I, Baba T, Yuzawa H, Kobayashi I, Cui L, Oguchi A, Aoki K, Nagai Y, Lian J, Ito T, Kanamori M, Matsumaru H, Maruyama A, Murakami H, Hosoyama A, Mizutani-Ui Y, Takahashi NK, Sawano T, Inoue R, Kaito C, Sekimizu K, Hirakawa H, Kuhara S, Goto S, Yabuzaki J, Kanehisa M, Yamashita A, Oshima K, Furuya K, Yoshino C, Shiba T, Hattori M, Ogasawara N, Hayashi H and Hiramatsu K (2001). Whole genome sequencing of meticillin-resistant *Staphylococcus aureus*. *Lancet*, **357**: 1225-1240.

23. Tang D. (2006) Using a supertree approach to detect laterally transferred genes within *Staphylococcus*. Brisbane: The University of Queensland. B.Sc. Thesis.

24. Hartl DL, Lozovskaya ER and Lawrence JG (1992). Nonautonomous transposable elements in prokaryotes and eukaryotes. *Genetica*, **86**: 47-53.

25. Inagaki Y, Susko E and Roger AJ (2006). Recombination between elongation factor 1-alpha genes from distantly related archaeal lineages. *Proc. Natl. Acad. Sci. U. S. A.*, **103**: 4528-4533.

26. Bork P and Doolittle RF (1992). Proposed acquisition of an animal protein domain by bacteria. *Proc. Natl. Acad. Sci. U. S. A.*, **89**: 8990-8994.

27. Omelchenko MV, Makarova KS, Wolf YI, Rogozin IB and Koonin EV (2003). Evolution of mosaic operons by horizontal gene transfer and gene displacement *in situ*. *Genome Biol.*, **4**: 55.

28. Igarashi N, Harada J, Nagashima S, Matsuura K, Shimada K and Nagashima KVP (2001). Horizontal transfer of the photosynthesis gene cluster and operon rearrangement in purple bacteria. *J. Mol. Evol.*, **52**: 333-341.

29. Chan CX, Beiko RG and Ragan MA (2007). Units of genetic transfer in prokaryotes. arXiv:0709.2027v1.

30. Baba T, Takeuchi F, Kuroda M, Yuzawa H, Aoki K, Oguchi A, Nagai Y, Iwama N, Asano K, Naimi T, Kuroda H, Cui L, Yamamoto K and Hiramatsu K (2002). Genome and virulence determinants of high virulence community-acquired MRSA. *Lancet*, **359**: 1819-27.

31. Holden MT, Feil EJ, Lindsay JA, Peacock SJ, Day NP, Enright MC, Foster TJ, Moore CE, Hurst L, Atkin R, Barron A, Bason N, Bentley SD, Chillingworth C, Chillingworth T, Churcher C, Clark L, Corton C, Cronin A, Doggett J, Dowd L, Feltwell T, Hance Z, Harris B, Hauser H, Holroyd S, Jagels K, James KD, Lennard N, Line A, Mayes R, Moule S, Mungall K, Ormond D, Quail MA, Rabbinowitsch E, Rutherford K, Sanders M, Sharp S, Simmonds M, Stevens K, Whitehead S, Barrell BG, Spratt BG and Parkhill J (2004). Complete genomes of two clinical *Staphylococcus aureus* strains: evidence for the rapid evolution of virulence and drug resistance. *Proc. Natl. Acad. Sci. U. S. A.*, **101**: 9786-91.

32. Diep BA, Gill SR, Chang RF, Phan TH, Chen JH, Davidson MG, Lin F, Lin J, Carleton HA, Mongodin EF, Sensabaugh GF and Perdreau-Remington F (2006). Complete genome sequence of USA300, an epidemic clone of community-acquired meticillin-resistant *Staphylococcus aureus*. *Lancet*, **367**: 731-9.

33. Iandolo JJ, Worrell V, Groicher KH, Qian Y, Tian R, Kenton S, Dorman A, Ji H, Lin S, Loh P, Qi S, Zhu H and Roe BA (2002). Comparative analysis of the genomes of the temperate bacteriophages phi 11, phi 12 and phi 13 of *Staphylococcus aureus* 8325. *Gene*, **289**: 109-118.

34. Herron LL, Chakravarty R, Dwan C, Fitzgerald JR, Musser JM, Retzel E and Kapur V (2002). Genome sequence survey identifies unique sequences and key virulence genes with unusual rates of amino Acid substitution in bovine Staphylococcus aureus. *Infect. Immun.*, **70**: 3978-3981.

35. Zhang YQ, Ren SX, Li HL, Wang YX, Fu G, Yang J, Qin ZQ, Miao YG, Wang WY, Chen RS, Shen Y, Chen Z, Yuan ZH, Zhao GP, Qu D, Danchin A and Wen YM (2003). Genome-based analysis of virulence genes in a non-biofilm-forming *Staphylococcus epidermidis* strain (ATCC 12228). *Mol. Microbiol.*, **49**: 1577-1593.

36. Takeuchi F, Watanabe S, Baba T, Yuzawa H, Ito T, Morimoto Y, Kuroda M, Cui L, Takahashi M, Ankai A, Baba S, Fukui S, Lee JC and Hiramatsu K (2005). Whole-genome sequencing of *staphylococcus haemolyticus* uncovers the extreme plasticity of its genome and the evolution of human-colonizing staphylococcal species. *J. Bacteriol.*, **187**: 7292-7308.

37. Kuroda M, Yamashita A, Hirakawa H, Kumano M, Morikawa K, Higashide M, Maruyama A, Inose Y, Matoba K, Toh H, Kuhara S, Hattori M and Ohta T (2005). Whole genome sequence of *Staphylococcus saprophyticus* reveals the pathogenesis of uncomplicated urinary tract infection. *Proc. Natl. Acad. Sci. U. S. A.*, **102**: 13272-13277.

38. Higgins CF (1992). ABC transporters: from microorganisms to man. *Annu. Rev. Cell Biol.*, **8**: 67-113.

39. Saier MH, Jr. and Reizer J (1994). The bacterial phosphotransferase system: new frontiers 30 years later. *Mol. Microbiol.*, **13**: 755-764.

40. Minin VN, Dorman KS, Fang F and Suchard MA (2005). Dual multiple change-point model leads to more accurate recombination detection. *Bioinformatics*, **21**: 3034-3042.

41. Huelsenbeck JP and Ronquist F (2001). MRBAYES: Bayesian inference of phylogenetic trees. *Bioinformatics*, **17**: 754-755.

42. Chan CX, Beiko RG and Ragan MA (2007). A two-phase strategy for detecting recombination in nucleotide sequences. *South African Computer Journal*, **38**: 20-27.

43. Bruen TC, Philippe H and Bryant D (2006). A simple and robust statistical test for detecting the presence of recombination. *Genetics*, **172**: 2665-2681.

44. Maynard Smith J (1992). Analyzing the mosaic structure of genes. *J. Mol. Evol.*, **34**: 126-129.

45. Jakobsen IB and Easteal S (1996). A program for calculating and displaying compatibility matrices as an aid in determining reticulate evolution in molecular sequences. *CABIOS*, **12**: 291-295.

46. Ragan MA (1992). Phylogenetic inference based on matrix representation of trees. *Mol. Phylogenet. Evol.*, **1**: 53-58.

47. Takahashi T, Satoh I and Kikuchi N (1999). Phylogenetic relationships of 38 taxa of the genus *Staphylococcus* based on 16S rRNA gene sequence analysis. *Int. J. Syst. Bacteriol.*, **49**: 725-728.

48. Finn RD, Mistry J, Schuster-Bockler B, Griffiths-Jones S, Hollich V, Lassmann T, Moxon S, Marshall M, Khanna A, Durbin R, Eddy SR, Sonnhammer EL and Bateman A (2006). Pfam: clans, web tools and services. *Nucleic Acids Res.*, **34**: D247-D251.

49. Andreeva A, Howorth D, Brenner SE, Hubbard TJ, Chothia C and Murzin AG (2004). SCOP database in 2004: refinements integrate structure and sequence family data. *Nucleic Acids Res.*, **32**: D226-D229.

50. Zhaxybayeva O, Gogarten JP, Charlebois RL, Doolittle WF and Papke RT (2006). Phylogenetic analyses of cyanobacterial genomes: quantification of horizontal gene transfer events. *Genome Res.*, **16**: 1099-1108.

51. Woese CR (2000). Interpreting the universal phylogenetic tree. *Proc. Natl. Acad. Sci. U. S. A.*, **97**: 8392-8396.

52. Lynch M and Conery JS (2000). The evolutionary fate and consequences of duplicate genes. *Science*, **290**: 1151-1155.

53. Hooper SD and Berg OG (2003). Duplication is more common among laterally transferred genes than among indigenous genes. *Genome Biol*, **4**: R48.

54. Chan CX, Beiko RG, Darling AE and Ragan MA (2007). Protein domains as units of genetic transfer. arXiv:0709.2030v1.

55. Archibald JM and Roger AJ (2002). Gene duplication and gene conversion shape the evolution of archaeal chaperonins. *J. Mol. Biol.*, **316**: 1041-1050.

56. Enright AJ, Van Dongen S and Ouzounis CA (2002). An efficient algorithm for large-scale detection of protein families. *Nucleic Acids Res.*, **30**: 1575-1584.

57. Notredame C, Higgins DG and Heringa J (2000). T-Coffee: a novel method for fast and accurate multiple sequence alignment. *J. Mol. Biol.*, **302**: 205-217.

58. Edgar RC (2004). MUSCLE: multiple sequence alignment with high accuracy and high throughput. *Nucleic Acids Res.*, **32**: 1792-1797.

59. Beiko RG, Chan CX and Ragan MA (2005). A word-oriented approach to alignment validation. *Bioinformatics*, **21**: 2230-2239.

60. Harlow TJ, Gogarten JP and Ragan MA (2004). A hybrid clustering approach to recognition of protein families in 114 microbial genomes. *BMC Bioinformatics*, **5**: 45.

61. Creevey CJ, Fitzpatrick DA, Philip GK, Kinsella RJ, O'Connell MJ, Pentony MM, Travers SA, Wilkinson M and McInerney JO (2004). Does a tree-like phylogeny only exist at the tips in the prokaryotes? *Proceedings of the Royal Society of London B: Biological Sciences*, **271**: 2551-2558.

62. Swofford DL. (2003) *PAUP*: Phylogenetic Analysis Using Parsimony (*and other methods)*. Sunderland MA.

63. Kimura M (1980). A simple method for estimating evolutionary rates of base substitutions through comparative studies of nucleotide sequences. *J. Mol. Evol.*, **16**: 111-120.

64. Shimodaira H and Hasegawa M (1999). Multiple comparisons of log-likelihoods with applications to phylogenetic inference. *Mol. Biol. Evol.*, **16**: 1114-1116.

65. Kishino H and Hasegawa M (1989). Evaluation of the maximum likelihood estimate of the evolutionary tree topologies from DNA sequence data, and the branching order in *Hominoidea*. *J. Mol. Evol.*, **29**: 170-179.

66. Goldman N, Anderson JP and Rodrigo AG (2000). Likelihood-based tests of topologies in phylogenetics. *Syst. Biol.*, **49**: 652-670.

67. Strimmer K and Rambaut A (2002). Inferring confidence sets of possibly misspecified gene trees. *Proceedings of the Royal Society of London B: Biological Sciences*, **269**: 137-142.

68. Strimmer K and von Haeseler A (1996). Quartet puzzling: A quartet maximum-likelihood method for reconstructing tree topologies. *Mol. Biol. Evol.*, **13**: 964-969.

69. Johnson NL, Kotz S and Kemp AW (1992). *Univariate Discrete Distributions*. 2nd ed. New York: Wiley.

70. Stoltzfus A, Spencer DF, Zuker M, Logsdon JM and Doolittle WF (1994). Testing the exon theory of genes: the evidence from protein structure. *Science*, **265**: 202-207.

71. Stoltzfus A, Spencer DF and Doolittle WF (1995). Methods for evaluating exon-protein correspondences. *CABIOS*, **11**: 509-515.

72. Massey FJ (1951). The Kolmogorov-Smirnov test for goodness of fit. *J. Am. Stat. Assoc.*, **46**: 68-78.